\documentclass[aps,prd,amsmath,amssymb,reprint,groupedaddress,superscriptaddress,nofootinbib]{revtex4-2}

\usepackage{graphicx}
\usepackage{dcolumn}
\usepackage{bm}
\usepackage{newtxtext,newtxmath}
\usepackage{hyperref}

\usepackage{xcolor}
\usepackage{soul}

\usepackage[T1]{fontenc}

\usepackage{amsmath}	
\usepackage{booktabs}

\newcommand{\lcdm}{$\Lambda$CDM}
\newcommand{\wcdm}{$w$CDM}
\newcommand{\wowacdm}{$w_0 w_a$CDM}
\newcommand{\prob}{\ensuremath{{p}}}
\newcommand{\data}{\ensuremath{\boldsymbol{{d}}}}
\newcommand{\evidence}{\ensuremath{z}}
\newcommand{\software}[1]{\footnote{\protect\url{#1}}}

\begin{document}

\title[Field-level cosmological model selection]{Field-level cosmological model selection: field-level simulation-based inference\\ for Stage IV cosmic shear can distinguish dynamical dark energy}

\author{Alessio Spurio Mancini}
\email{alessio.spuriomancini@rhul.ac.uk}
\affiliation{Department of Physics, Royal Holloway, University of London, Egham Hill, Egham, UK}

\author{Kiyam Lin}%
\affiliation{Mullard Space Science Laboratory, University College London, Holmbury St.\ Mary, Dorking, Surrey, RH5 6NT, UK}

\author{Jason D. McEwen}%
\affiliation{Mullard Space Science Laboratory, University College London, Holmbury St.\ Mary, Dorking, Surrey, RH5 6NT, UK}

\date{\today}

\begin{abstract}
	We present a framework that for the first time allows Bayesian model comparison to be performed for field-level inference of cosmological models.  We achieve this by taking a simulation-based inference (SBI) approach using neural likelihood estimation, which we couple with the learned harmonic mean estimator in order to compute the Bayesian evidence for model comparison.  We apply our framework to mock Stage IV cosmic shear observations to assess its effectiveness at distinguishing between various models of dark energy.  If the recent DESI results that provided exciting hints of dynamical dark energy were indeed the true underlying model, our analysis shows Stage IV cosmic shear surveys could definitively detect dynamical dark energy.  We also perform traditional power spectrum likelihood-based inference for comparison, which we find is not able to distinguish between dark energy models, highlighting the enhanced constraining power for model comparison of our field-level SBI approach.
\end{abstract}

\maketitle


\section{\label{sec:introduction}Introduction}
Central questions in cosmology are often those of model comparison. For example, what model best describes the underlying nature of dark energy?  The concordance \lcdm\ model attributes dark energy to Einstein’s cosmological constant $\Lambda$. In the \wcdm\ model, the dark energy equation-of-state parameter $w$ is constant in time but allowed to vary from -1 to be constrained by observations. In dynamical dark energy models, such as the \wowacdm\ model, $w$ is free to evolve over time.
Recent results from the DESI collaboration \citep{adame2024desi} hint at the exciting possibility of dynamical dark energy, although the consensus at present is that there is no evidence to prefer a model more complicated than \lcdm.
In this work we present a cosmic shear analysis pipeline for cosmological model comparison and study its effectiveness at comparing the \lcdm, \wcdm\ and \wowacdm\ cosmological models.

While traditional cosmological inference pipelines are typically based on likelihood-based analysis of two-point statistics, it is widely known that probes of the large-scale structure contain a great deal of cosmological information beyond two-point statistics due to the non-linear nature of gravity.
Field-level inference is capable of capturing this higher-order statistical information \citep[\textit{e.g.}][]{boruah2022map, tsaprazi2022field, porqueres2022lifting, andrews2023bayesian, porqueres2023field, lanzieri2024optimal, Nguyen24}).
Upcoming Stage IV surveys of the large-scale structure, such as \emph{Euclid} \citep{laureijs2011euclid}, Rubin Observatory Legacy Survey of Space and Time (Rubin-LSST) \citep{lsst2009lsst} or \emph{Roman} \citep{spergel2015wide} will acquire data that contain significant high-order cosmological information in the observed fields.

However, field-level inference with such data is challenging due to the high-dimensional nature of the parameter space to be inferred and the complexity of the forward model. Typically only parameter estimation is considered, as model comparison is too computationally costly.
Yet, the more information we acquire, the better we can distinguish between underlying models. Consequently, performing model comparison on field-level data may enable us to definitively determine which dark energy model best describes our Universe.

An alternative to the aforementioned likelihood-based field-level inference approaches are those that employ simulation-based inference (SBI) \citep[\textit{e.g.} ][]{cranmer2020frontier}.
In this paradigm it is possible to run forward simulations that are able to fully propagate all known uncertainties from parameters to data without needing to explicitly define their corresponding probability distributions.
Thus, this approach captures all uncertainties in the data without any statistical simplifications.
Modern SBI approaches based on neural density estimation have been applied successfully to two-point statistical analyses in cosmology \citep{alsing2019fast, taylor2019cosmic, lin2023simulation, mvwk2024kids} and to field-level analyses \citep{jeffrey2021likelihood, lemos2023simbig, gatti2024dark, jeffrey2024dark, lin2024scattering}. Since SBI methods still require large numbers of simulations for training, accelerating simulations is important and becomes even more pertinent for field-level analyses.  In particular, neural emulators, such as \texttt{CosmoPower} \citep{CP}, can offer considerable computational savings that in many cases are essential. Furthermore, they also typically support automatic differentiation, which can be leveraged for further acceleration or to reduce the volume of training data needed \citep{brehmer2020mining, zeghal2022neural}.

Bayesian model comparison provides a principled framework to distinguish between models---naturally incorporating Occam's razor to trade off model complexity and goodness of fit---that has already found widespread use in cosmology \citep{Trotta07}. Model comparison requires computation of the Bayesian evidence, which is computationally challenging even in moderate numbers of dimensions.  Nested sampling \citep{Skilling06} is often used to compute the evidence, as implemented in numerous algorithms \citep[\textit{e.g.}][]{Feroz08, Handley15, Handley15b, Feroz19, higson2019dynamic, speagle2020dynesty, cai2022proximal, lange2023nautilus}, although this requires coupling sampling and evidence calculation.  Recently, the learned harmonic mean estimator \citep{mcewen2021machine} has been presented as an alternative to nested sampling that is flexible, robust, and scalable \citep{Polanska23,polanska2024learned,piras2024future}.  Moreover, as the learned harmonic mean requires posterior samples only, it is agnostic to the sampling strategy adopted and so can be combined with accelerated sampling techniques, such as the No U-Turn Sampler (NUTS; \cite{Hoffman14}), \texttt{FlowMC} \citep{Wong22}, or others, as demonstrated already \citep{piras2024future, polanska2024flowmc}.
However, Bayesian model comparison has not typically been considered for field-level analyses due to the high-dimensional parameter spaces involved and the difficulties in scaling evidence computation to those dimensions (a notable exception is proximal nested sampling \cite{cai2022proximal,mcewen2023proximal}, which has been scaled to field-level inference but that is restricted to convex likelihoods and so not applicable to complex cosmological models).

While field-level SBI approaches capture high-order statistical information from the field, the underlying parameter space includes parameters of interest only.  In contrast, likelihood-based field-level approaches consider the pixels of the observed or initial field as parameters to be inferred, resulting in very high dimensional parameter spaces.  The reduced parameter dimension of field-level SBI opens up the possibility of cosmological model comparison for field-level inference.  Model comparison for modern neural SBI approaches was first considered by Ref. \cite{spurio2023bayesian}, where the flexibility of the learned harmonic mean estimator was exploited.
Alternative approaches to estimate the evidence that are applicable for SBI have since been introduced where a model is trained specifically to compute the evidence \citep{jeffrey2024evidence,radev2021amortized}.

In this work we present a framework for field-level Bayesian inference, where for the first time we consider not only parameter estimation but also model selection.  This is achieved by performing field-level SBI, specifically neural likelihood estimation (NLE;  \cite{papamakarios2019sequential}), where cosmological forward models are accelerated by the \texttt{CosmoPower} emulator \citep{CP,CPJ}, that we couple with the learned harmonic mean for cosmological model comparison \citep{mcewen2021machine,polanska2024learned,spurio2023bayesian}.
We demonstrate a field-level pipeline on simulated cosmic shear observations, showing that Stage IV surveys can distinguish between different models of dark energy. For comparison, we also consider a likelihood-based analysis based on two-point statistics and demonstrate that it is not able to distinguish between different models, emphasising the effectiveness of field-level inference.

The remainder of this article is structured as follows. Section \ref{sec:methods} details the methodology introduced for field-level SBI that also supports cosmological model comparison. In Section~\ref{sec:results} we apply our framework to mock Stage IV cosmic shear observations to assess its effectiveness at distinguishing between various models of dark energy. Concluding remarks are made in Section~\ref{sec:conclusions}.

\section{\label{sec:methods}Methodology}

We introduce a framework for field-level inference that also supports cosmological model comparison.  Specifically, we consider an SBI approach based on NLE (neural likelihood estimation) that we couple with the learned harmonic mean estimator.  First, however, we outline the traditional power spectrum likelihood-based inference approach that we consider for comparison.

\subsection{Power spectrum likelihood-based inference} \label{sec:methods_ps}

For comparison purposes, we perform a likelihood-based analysis of the weak lensing shear power spectrum. The likelihood is assumed to be Gaussian, following the setup presented in Ref. \cite{lanzieri2024optimal}. The log-likelihood is given by
\begin{equation}
	\log \prob(\bm{d} \vert \bm{\theta}) = -\frac{1}{2}[\bm{d}-\bm{\mu}(\bm{\theta})]^{T}\bm{C}^{-1}[\bm{d}-\bm{\mu}(\bm{\theta})],
\end{equation}
up to a constant, where $\bm{\theta}$ represents the underlying cosmological parameters, $\bm{d}$ is the data vector and $\bm{C}$ is the covariance matrix for a fixed fiducial cosmology (the same used to generate the mock data vector). The theory shear power spectrum $\bm{\mu}(\bm{\theta})$ is calculated from the underlying matter power spectrum using \href{https://github.com/DifferentiableUniverseInitiative/jax\_cosmo}{\texttt{jax-cosmo}}\software{https://github.com/DifferentiableUniverseInitiative/jax\_cosmo} \citep{campagne2023jax}.
The non-linear matter power spectrum is provided by \href{https://github.com/dpiras/cosmopower-jax}{\texttt{CosmoPower-JAX}}\software{https://github.com/dpiras/cosmopower-jax} \citep{CPJ}, a JAX implementation of  \href{https://github.com/alessiospuriomancini/cosmopower}{\texttt{CosmoPower}} \citep{CP}, which provides a neural network to emulate the non-linear matter power spectrum. In this work we couple \texttt{CosmoPower-JAX} with \texttt{jax-cosmo}, using the former to emulate the non-linear prescription given by \texttt{HMCode} \citep{mead2016accurate}. \texttt{HMCode} provides a parameterised prescription to account for baryonic feedback; in this analysis we fix the baryonic parameters $c_{\rm min}$ and $\eta_0$ to their dark matter-only values, 3.13 and 0.603, respectively.

For the simulated data, we make use of a modified version of \href{https://github.com/DifferentiableUniverseInitiative/sbi_lens}{\texttt{sbi\_lens}}\software{https://github.com/DifferentiableUniverseInitiative/sbi\_lens} \citep{lanzieri2024optimal} (adding support for the \wowacdm\ model) to generate correlated convergence maps following a log-normal prescription with Gaussian noise across five tomographic bins.  The simulated data is configured to approximately mimic a Stage IV survey.  We use \href{https://github.com/apetri/LensTools}{\texttt{lenstool}}\software{https://github.com/apetri/LensTools} \citep{petri2016lens} to calculate the auto and cross power spectra from the simulated noisy convergence maps.

To generate posterior samples we perform Markov chain Monte Carlo (MCMC) sampling using the NUTS \citep{Hoffman14} sampler implemented in the \href{https://github.com/pyro-ppl/numpyro}{\texttt{NumPyro}}\software{https://github.com/pyro-ppl/numpyro} differentiable probabilistic programming library \citep{phan2019composable, bingham2019pyro}.

\subsection{Field-level SBI inference} \label{sec:methods_field}

To perform field-level SBI we do not need an analytical prescription of the likelihood. Instead, we take an NLE (neural likelihood estimation; \cite{papamakarios2019sequential}) approach and learn an implicit likelihood from forward simulated data-parameter pairs.
In contrast to Ref. \cite{lanzieri2024optimal} who adopt neural posterior estimation (NPE; \cite{Papamakarios16, lueckmann2017flexible, Greenberg19}), we adopt NLE since it provides greater flexibility in the choice of proposal used for generating training data and the neural density estimator can be integrated within an MCMC framework that provides statistical guarantees.  Moreover, it simplifies evidence calculation with the learned harmonic mean \citep{spurio2023bayesian}.\footnote{The learned harmonic mean requires the evaluation of the likelihood, or its surrogate, at posterior samples, hence model comparison with NPE requires two density estimator to be trained \citep{spurio2023bayesian}.}

NLE involves training a conditional density estimator $q_{\boldsymbol{\phi}} (\data \vert \boldsymbol{\theta})$ to act as a surrogate for the likelihood $\prob(\data \vert \boldsymbol{\theta})$ (considering it as a probability distribution over the data), where $\bm{\phi}$ represent the parameters of the density estimator (\text{i.e.} neural network weights).   Given paired training data $\{ \bm{\theta}_i, \bm{d}_i\}$ for parameters drawn from an arbitray proposal $\bm{\theta}_i \sim \tilde{p}(\bm{\theta})$, the NLE density estimator can be trained by minimising the negative log-likelihood of the surrogate, which can be shown to be equivalent to maximising the Kullback-Leibler divergence $D_\text{KL}(\cdot \| \cdot)$ between the likelihood and its surrogate: \begin{align}
	\mathbb{E}_{\prob(\data \vert\boldsymbol{\theta}) \tilde{\prob}(\boldsymbol{\theta})}\bigl[ -\text{log}\,q_{\boldsymbol{\phi}}(\data \vert \boldsymbol{\theta}) \bigr]
	= \mathbb{E}_{\tilde{\prob}(\theta)} \bigl[ D_\text{KL}( \prob(\data \vert \bm{\theta}) \|  q_{\bm{\phi}}(\data \vert \bm{\theta}) \bigr],
\end{align}
up to a constant. Consequently, when trained in this manner the density estimator learns to approximate the likelihood over the parameter space covered by the proposal distribution.
For this work we use normalizing flows \citep{papamakarios2021normalizing} as the conditional neural density estimator.  Specifically, we adopt a masked autoregressive flow (MAF; \cite{papamakarios2017masked}) constructed out of masked autoencoders for density estimation (MADE; \cite{germain2015made}).
In terms of implementation, we make use of the \href{https://github.com/sbi-dev/sbi}{\texttt{sbi}}\software{https://github.com/sbi-dev/sbi} software package  \citep{tejero-cantero2020sbi} to construct and train the NLE density estimator.

We follow the same simulation procedure as described in Section~\ref{sec:methods_ps}.
Specifically, we make use of \href{https://github.com/dpiras/cosmopower-jax}{\texttt{CosmoPower-JAX}} to simulate the matter power spectrum, \href{https://github.com/DifferentiableUniverseInitiative/jax\_cosmo}{\texttt{jax-cosmo}} to compute the shear power spectrum, and a modified version of \href{https://github.com/DifferentiableUniverseInitiative/sbi_lens}{\texttt{sbi\_lens}} to then generate correlated log-normal convergence maps with Gaussian noise.

While modern neural density estimators can scale to relatively high dimensional settings, for field-level SBI it is typical to compress the field to a lower dimensional latent representation, for example by neural, statistical or wavelet scattering based compression techniques \citep[\textit{e.g.}][]{tegmark1997karhunen, heavens2000massive, alsing2018generalized, charnock2018imnn, jeffrey2021likelihood, makinen2023fishnets, gatti2024dark}. While wavelet scattering transforms have recently been shown to be effective for this purpose and do not require additional simulations \citep{lin2024scattering}, for the purposes of this work we consider neural compression.
An extensive study of neural compression techniques for SBI was recently performed by Ref. \cite{lanzieri2024optimal}, demonstrating that a convolutional neural network (CNN) trained with a variational mutual information maximisation (VMIM; \cite{jeffrey2021likelihood}) loss function can achieve excellent compression performance, capturing close to all higher order cosmological information in cosmic shear fields. We therefore adopt this neural compression technique and make use of the ResNet-18 CNN architecture \citep{he2016deep} implemented in \href{https://github.com/google-deepmind/dm-haiku}{\texttt{Haiku}}\software{https://github.com/google-deepmind/dm-haiku} \citep{hennigan2020haiku} to compress the convergence maps, training our own compressor following the same procedure described in \citep{zeghal2024simulation} and included in \href{https://github.com/DifferentiableUniverseInitiative/sbi_lens}{\texttt{sbi\_lens}}.

Finally, posterior samples are then generated by MCMC sampling using the surrogate likelihood.  In this case, for simplicity we use the \href{https://github.com/dfm/emcee}{\texttt{emcee}}\software{https://github.com/dfm/emcee} software package \citep{foreman2013emcee} to perform sampling, although alternative accelerated sampling techniques could be considered.

Our overall pipeline, including the simulator, is automatically differentiable, which can provide numerous advantages.  While an automatically differentiable simulator is not strictly necessary for the SBI results presented in this work, it could in principle be used to train the NLE model more efficiently, requiring less training data \citep{brehmer2020mining, zeghal2022neural}.  However, Ref. \cite{zeghal2024simulation} recently found that incorporating gradients provided by automatic differentiation did not significantly improve field-level SBI inference and so we have not considered this further in the current article. Alternatively, the differentiable forward model is essential for field-level likelihood-based inference with a Bayesian hierarchical model (BHM) in order to leverage high-dimensional sampling techniques that exploit gradient information (\textit{e.g.} NUTS).  However, such an approach does not at present support Bayesian model selection and so we have not considered it further in the current article.  We will present a field-level BHM approach that can also provide the calculation of the Bayesian evidence in an upcoming work.

\subsection{\label{sec:methods_evidence}Bayesian evidence for model comparison}

For both settings considered previously, namely for both power spectrum likelihood-based inference and field-level SBI inference, we recover posterior samples by MCMC sampling.  Moreover, for each sample the unnormalized posterior density will be evaluated during sampling.  Thus, we have access to everything needed to compute the Bayesian evidence using the learned harmonic mean estimator, irrespective of the underlying method used to generate the posterior samples.

The Bayesian evidence is given by the marginalised likelihood
\begin{align}\label{eq:evidence}
	\evidence = \prob ( \data \vert M ) = \int \mathrm{d} \boldsymbol{\theta} \, \prob ( \data \vert \boldsymbol{\theta}, M ) \prob ( \boldsymbol{\theta} \vert M ),
\end{align}
for likelihood $\prob ( \data \vert \boldsymbol{\theta}, M )$ and prior $\prob ( \boldsymbol{\mathrm{\theta}} \vert M )$, where here we have made the model $M$ explicit.
The evidence is a critical term to compute in order to compare models. The posterior model odds between two competing models $M_1$ and $M_2$ can be written as
\begin{align}
	\frac{\prob ( M_1 \vert \data )}{\prob ( M_2 \vert \data )} = \frac{\prob ( \data  \vert M_1 ) \prob ( M_1 )}{\prob ( \data  \vert M_2 ) \prob ( M_2 )},
\end{align}
which follows by Bayes' theorem.  In many cases \textit{a priori} probabilities $\prob (M_1)$ and $\prob (M_2)$ of the two models are considered to be equal, hence the ratio of posterior distributions becomes equivalent to the evidence ratio or Bayes factor
\begin{align}\label{eq:bayes_factor}
	B_{12} = \frac{\prob ( \data  \vert M_1 )}{\prob ( \data  \vert M_2 )} = \frac{\evidence_1}{\evidence_2}.
\end{align}
For notational brevity, henceforth we drop the explicit conditioning on models unless there are multiple models under consideration.

The learned harmonic mean can be used to compute the evidence for different models, and thus to also compute Bayes factors for Bayesian model comparison.  While the original harmonic mean \citep{newton1994approximate} suffered from an exploding variance \citep{neal:1994}, the learned harmonic mean solves this issue by integrating machine learning to learn an internal target distribution \citep{mcewen2021machine}.  Critically, the learned internal target must be concentrated within the posterior.  Normalizing flows provide an elegant way to ensure this simply by lowering the temperature $T$ (\textit{i.e.} variance) of their base distribution \citep{polanska2024learned}, avoiding the need for bespoke training. Given the learned target distribution $\varphi_{\bm{\psi}}(\bm{\theta}; T)$ with parameters $\bm{\psi}$ (\text{i.e.} neural network weights), the reciprocal evidence $\rho = \evidence^{-1}$ is then estimated as
\begin{equation}
	\label{eqn:harmonic_mean_retargeted}
	\hat{\rho} =
	\frac{1}{N} \sum_{i=1}^N
	\frac{\varphi_{\bm{\phi}}(\theta_i; T)}{\prob(\data \vert \bm{\theta}_i, M) \prob(\bm{\theta}_i \vert M)} ,
	\quad
	\bm{\theta}_i \sim \prob(\bm{\theta} \vert \data, M).
\end{equation}
We compute evidence estimates from posterior samples for both likelihood-based and SBI settings using the \href{https://github.com/astro-informatics/harmonic}{\texttt{harmonic}}\software{https://github.com/astro-informatics/harmonic} software package implementing the learned harmonic mean.

It is important to note that evidence values are of course sensitive to the choice of priors.  This is a feature of Bayesian model comparison and not a bug, as it encapsulates Occam's razor \citep{ghahramani2013bayesiany}.
In the Bayesian formalism models are specified as probability distributions over datasets and, since probability distributions must be normalized, each model has a limited ``probability budget'' to allocate. While a complex model can represent a wide range of datasets well, it spreads its predictive probability widely. In doing so, the model evidence of complex models will be penalised if such complexity is not required.
There are a wide variety of ways to set priors appropriately for Bayesian inference depending on the statistical question at hand \citep{llorente2023safe}.  For example, approaches to setting priors include physical priors (\textit{e.g.} non-negative mass or flux; \cite{carrillo2014purify}), uninformative Jeffreys priors that are invariant to a parameter transformation \citep{lee1989bayesian}, informative priors for example to regularize inverse problems \citep[\textit{e.g.}][]{price2021sparse}, data-driven priors potentially specified by a generative model \citep[\textit{e.g.}][]{mcewen2023proximal,remy2023probabilistic, liaudat2024scalable}, or data-informed priors, where the posterior of an \textit{a priori} analysis is used as the prior for an analysis with new data \citep[\textit{e.g.}][]{alsing2021nested}.
Despite a variety of methods to set appropriate piors, there remains debate regarding sensitivity of the evidence to prior choice \citep{linder2007tainted, liddle2007comment, efstathiou2008limitations}.  If one wishes to remove the prior dependence for the purpose of studying tensions between data-sets, once the evidence is estimated it can be used to compute the Bayesian suspiciousness \citep{handley2019quantifying, lemos2020quantifying}.

\section{\label{sec:results}Results}
We apply the power spectrum likelihood-based and field-level SBI inference frameworks outlined previously to simulated cosmic shear observatives intended to mimic a Stage IV survey.  We run MCMC sampling to generate posterior samples, which we also use for evidence estimation with the learned harmonic mean estimator to compare \wcdm\ and \wowacdm~models to \lcdm.
We present marginal posterior distributions of the cosmological parameters and Bayes factors for comparisons between the models considered, for a variety of ground truth data vectors.

While we follow the general methodology outlined in Section~\ref{sec:methods}, specific details of the simulator, neural compressor,  neural density estimator, MCMC samplers, and learned harmonic mean estimator can be found in Appendix~\ref{app:config}.

\subsection{Models, mock data \& priors}
We consider three models, the ubiquitous \lcdm~and \wcdm~cosmological models and also a phenomenological model that allows the equation of state for dark energy to evolve.
For this dynamical dark energy we adopt the Chevallier-Polarski-Linder (CPL) parameterisation with $w(a) = w_0 + w_a (1-a)$ \citep{chevallier2001accelerating, linder2003exploring}, where $a$ denotes the scale factor, resulting in the so-called \wowacdm\ model. For model comparison, we focus on comparing \wcdm\ and \wowacdm~models to \lcdm.  Recent results presented by the DESI collaboration \citep{adame2024desi} provide exciting hints of \wowacdm, although this is only for certain data combinations and thus is far from conclusive.

For the two model comparisons performed (\lcdm\ vs \wcdm\ and \lcdm\ vs \wowacdm) we consider two different ground truth mock data cases, one generated by each model, resulting in four total model comparisons.  Mock data cosmological parameter values and prior ranges are shown in Table~\ref{tab:param_priors}.
In our analysis for all parameters besides $(w_0,w_a)$ we consider the same priors as Ref.~\cite{zhang2022transitioning}, matching the Rubin-LSST science requirements document.  The ground truth is set to the middle of the prior range.
We set $w=-1$ in the \lcdm\ case.
For other models we consider fiducial parameters as if the DESI results hinting at dynamical dark energy were the ground truth.  That is, we set the $(w_0,w_a)$ ground truth to the best-fit DESI parameters for the data combination showing hints of dynamical dark energy (DESI $+$ CMB $+$ PantheonPlus data) \citep{adame2024desi} for both the \wcdm\ (with $w_0 = w$) and \wowacdm\ cases.
For $(w_0,w_a)$, the posterior distributions of the Dark Energy Survey (DES; \cite{dark2005dark}) year three data \citep{Abbott_2023} are used as the prior, \textit{i.e.} we follow a data-informed prior approach (see Section~\ref{sec:methods_evidence}).

\begin{table}
	\caption{Cosmological parameter values for the mock data and prior ranges. The normal distribution is denoted $\mathcal{N}$, while a truncated normal is denoted $\mathcal{N}_T$.  The distribution for $\Omega_{\text{cdm}}$ is truncated to have a lower bound of -1. The distribution for $w$ is truncated to have a lower bound of -2.0 and an upper bound of -0.33.}
	\centering
	\begin{tabular}{l c c}
		\toprule
		Parameter               & Mock data & Prior range                  \\
		\midrule
		$\Omega_{\mathrm{cdm}}$ & 0.2664    & $\mathcal{N}_T(0.2664, 0.2)$ \\
		$\Omega_{\mathrm{b}}$   & 0.0492    & $\mathcal{N}(0.0492, 0.006)$ \\
		$\sigma_8$              & 0.831     & $\mathcal{N}(0.831, 0.14)$   \\
		$h$                     & 0.6727    & $\mathcal{N}(0.6727, 0.063)$ \\
		$n_{\mathrm{s}}$        & 0.9645    & $\mathcal{N}(0.9645, 0.08)$  \\
		\midrule
		$w$ (\lcdm)             & -1.0      & -                            \\
		\midrule
		$w$ (\wcdm)             & -0.827    & $\mathcal{N}_T(-1.0, 0.9)$   \\
		\midrule
		$w_0$ (\wowacdm)        & -0.827    & $\mathcal{N}(-0.95, 0.08)$   \\
		$w_a$ (\wowacdm)        & -0.75     & $\mathcal{N}(-0.4, 0.4)$     \\
		\bottomrule
	\end{tabular}
	\label{tab:param_priors}
\end{table}

\subsection{\lcdm\ vs \wcdm}


\begin{figure*}
	\includegraphics[width=\columnwidth]{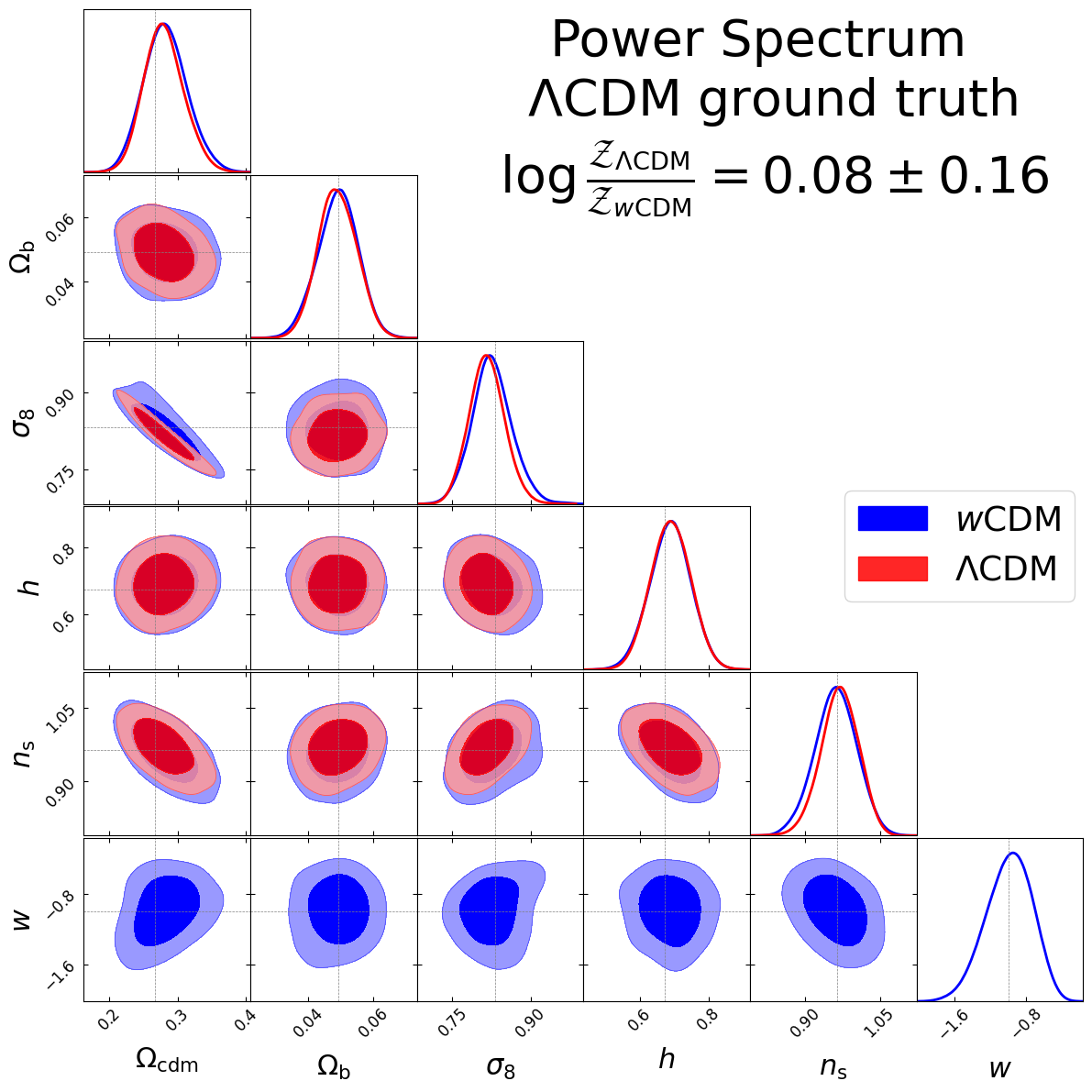}
	\includegraphics[width=\columnwidth]{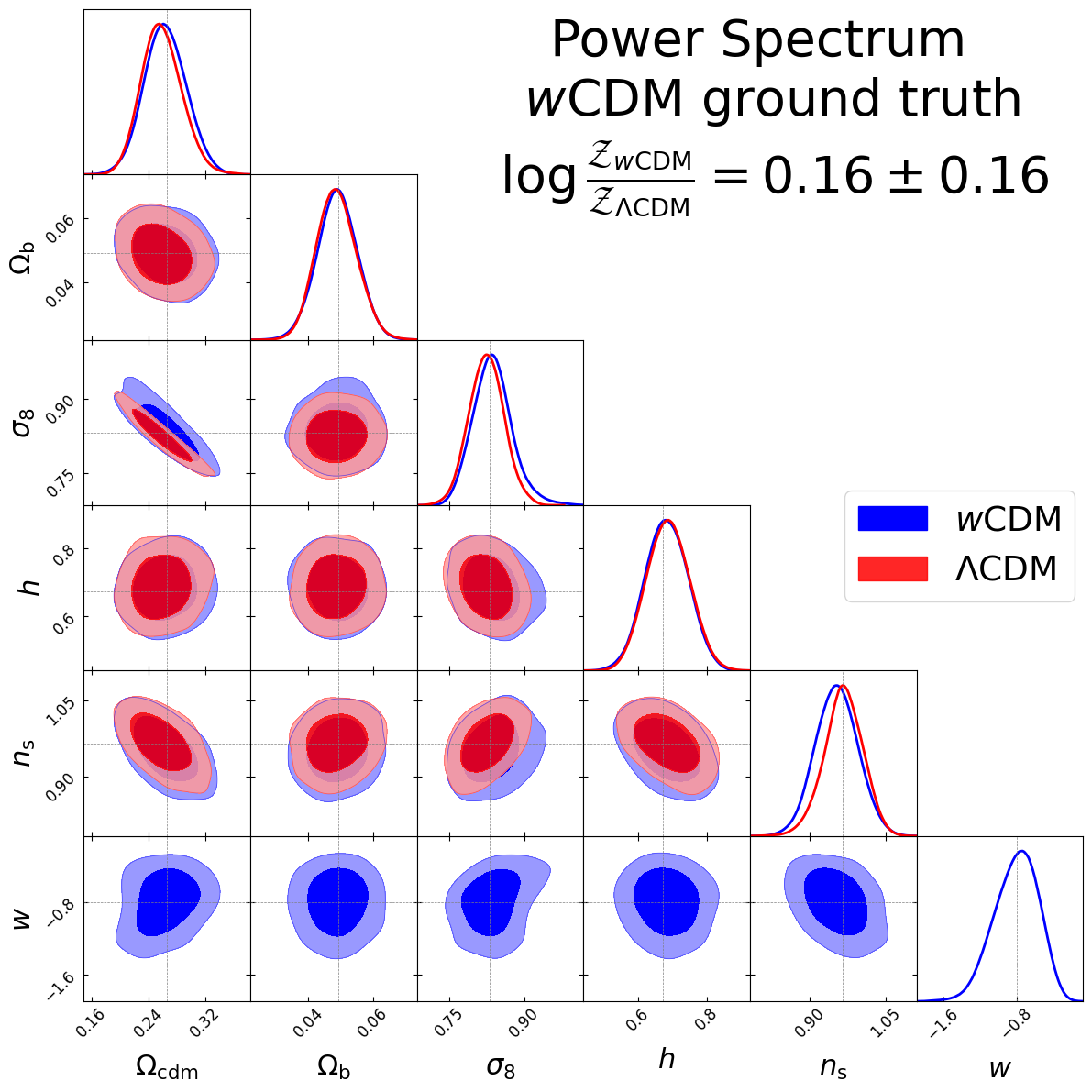}
	\caption{Marginal posterior distributions of cosmological parameters for the \textbf{power spectrum likelihood-based inference}, comparing \textbf{\lcdm\ vs \wcdm}. Ground truth underlying parameter values are indicated by dashed lines. Left: \lcdm\ ground truth data vector. Right: \wcdm\ ground truth data vector. For both ground truth scenarios the Bayesian evidence values show it is \textbf{not possibile to distinguish cosmological models}. }
	\label{fig:ps_w}
\end{figure*}

\begin{figure*}
	\includegraphics[width=\columnwidth]{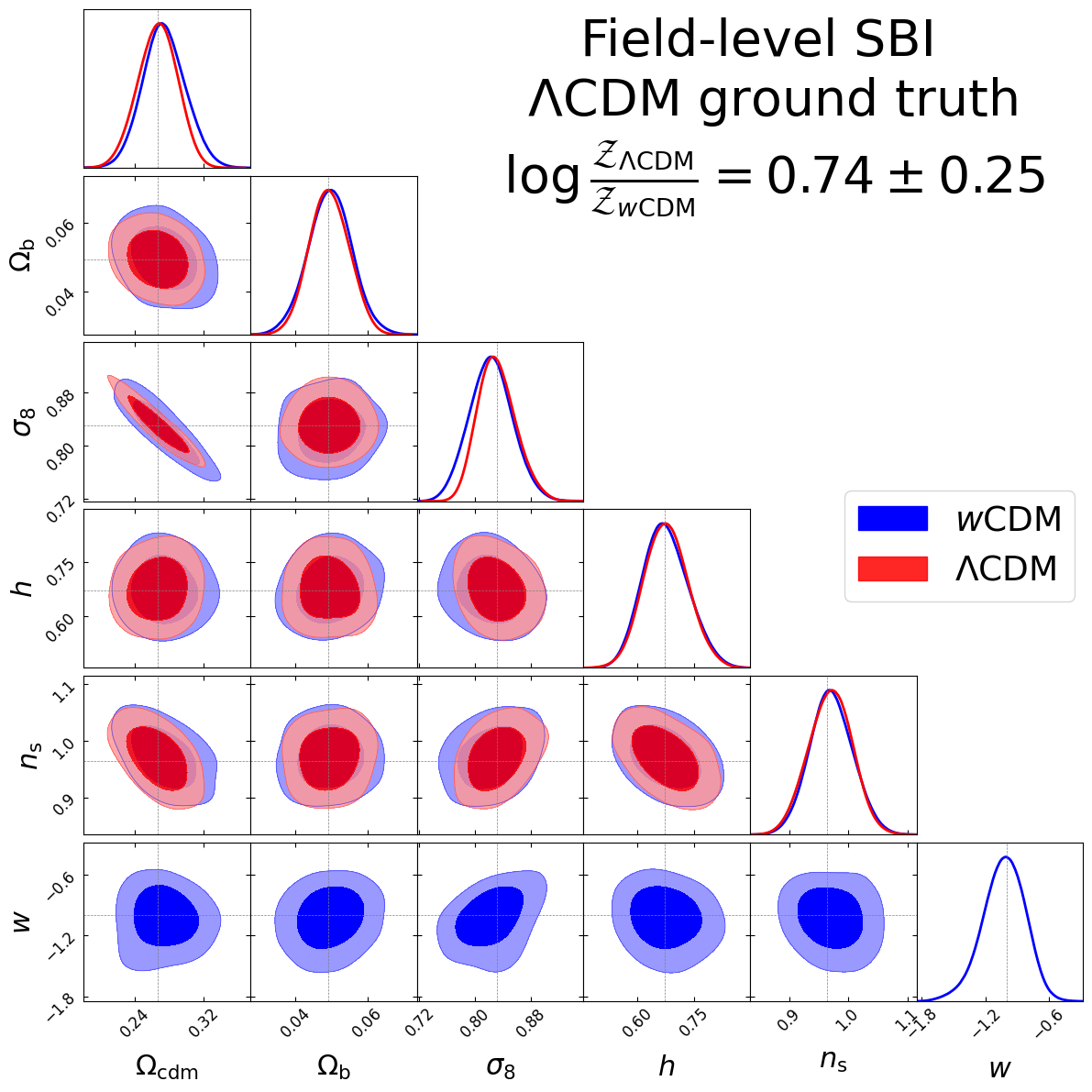}
	\includegraphics[width=\columnwidth]{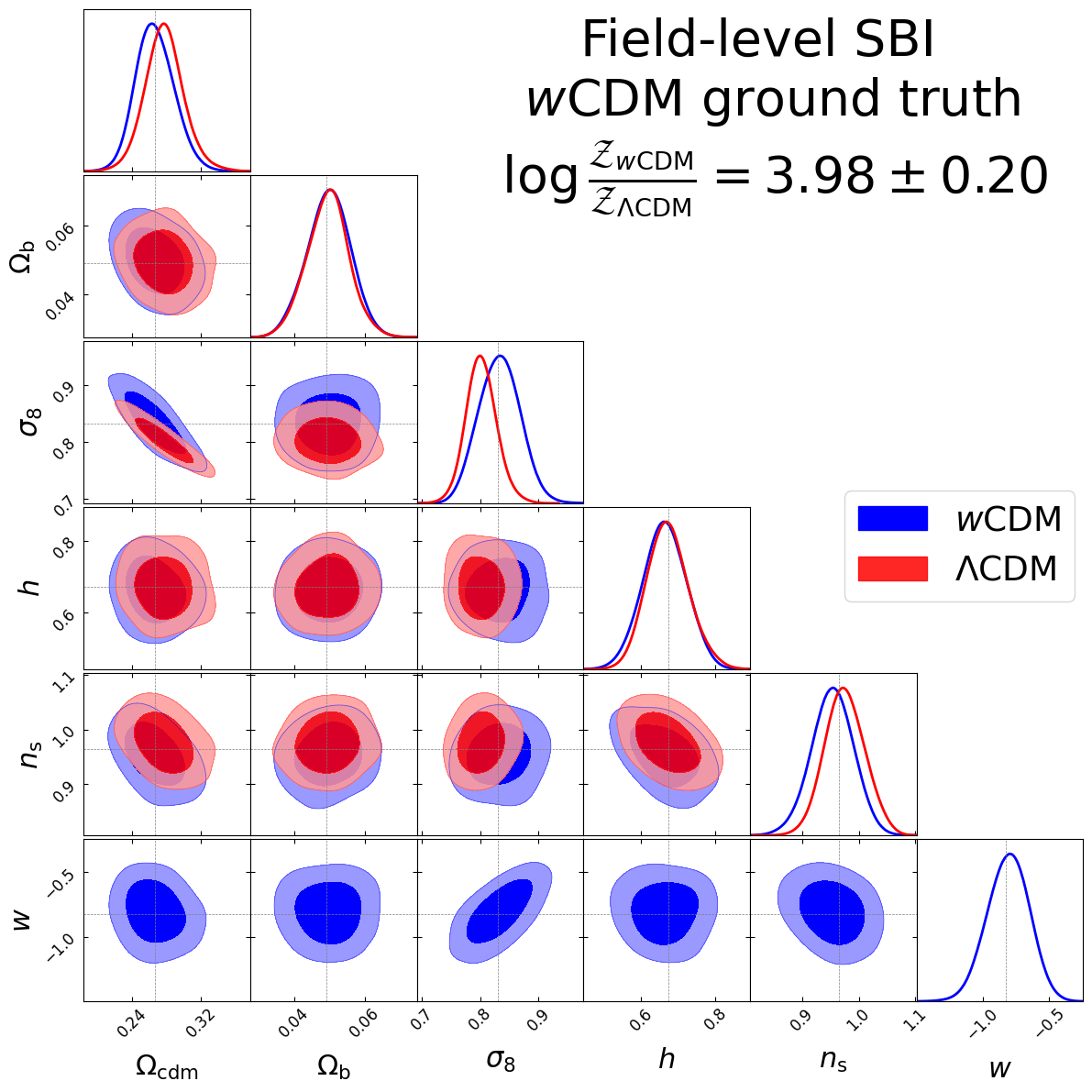}
	\caption{Marginal posterior distributions of cosmological parameters for the \textbf{field-level SBI inference}, comparing \textbf{\lcdm\ vs \wcdm}. Ground truth underlying parameter values are indicated by dashed lines. Left: \lcdm\ ground truth data vector. Right: \wcdm\ ground truth data vector. For the former ground truth scenario the Bayesian evidence \textbf{weakly prefers the true underlying model \lcdm}. For the latter ground truth scenario the Bayesian evidence \textbf{strongly pefers the true underlying model \wcdm}. }
	\label{fig:sbi_w}
\end{figure*}

\begin{figure}
	\includegraphics[width=\columnwidth]{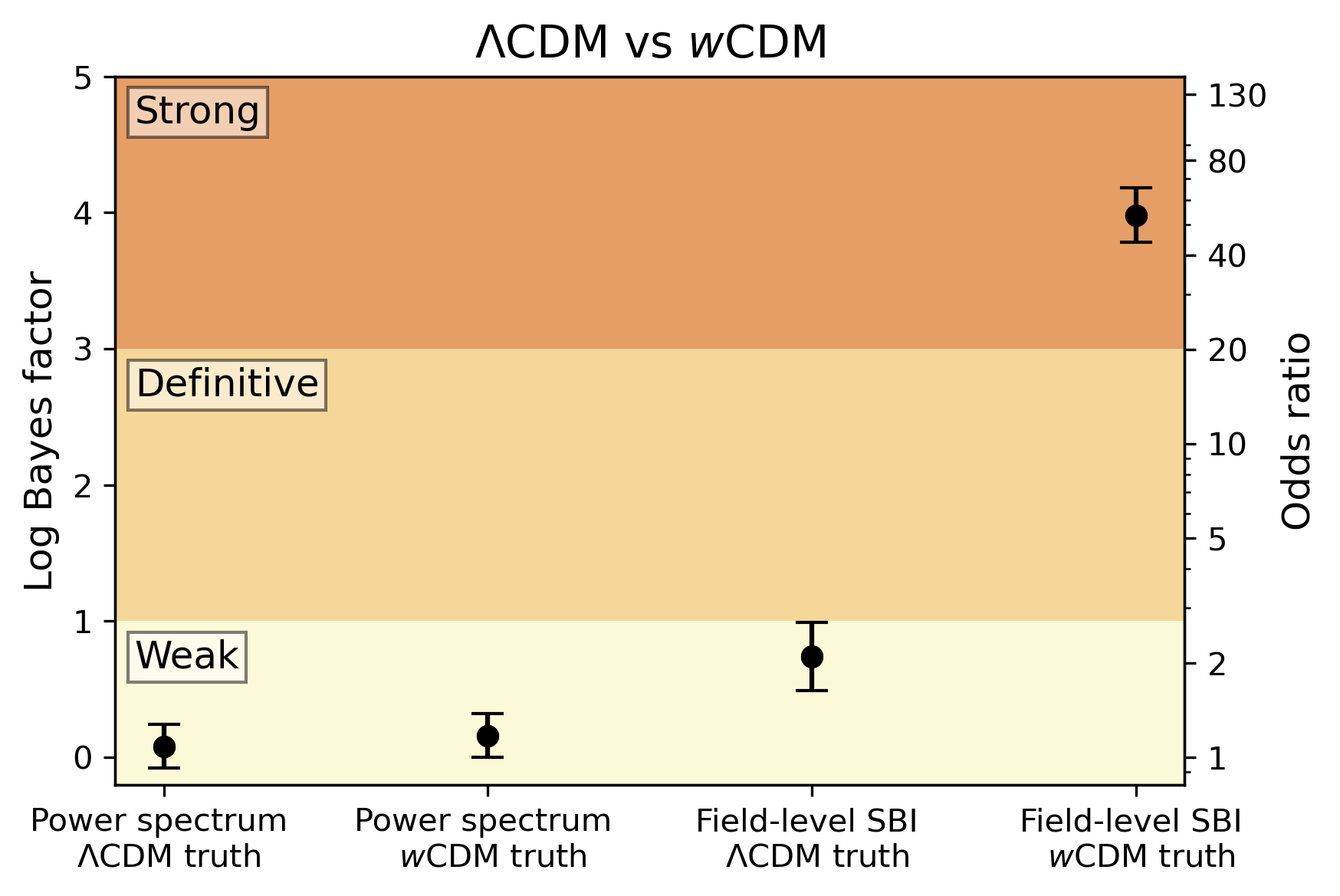}
	\caption{Bayes factors with errors for the \lcdm\ vs \wcdm\ comparison. The shaded regions correspond to the strength of the Bayes factor on the Jeffreys scale.  Note that \textbf{power spectrum likelihood-based inference cannot distinguish between \lcdm\ and \wcdm, whereas field-level SBI inference can}. }
	\label{fig:bayes_w}
\end{figure}

Figure~\ref{fig:ps_w} and Figure~\ref{fig:sbi_w} show marginalised posterior distributions recovered for the power spectrum likelihood-based inference and field-level SBI inference, respectively, for \lcdm\ vs \wcdm. Both figures show results for the two different ground truth mock data cases.
Bayes factors for each setting are displayed on each marginal distribution plot.  Furthermore, they are summarised visually in Figure~\ref{fig:bayes_w} for this \lcdm\ vs \wcdm\ comparison.
Bayes factors for all experiments are summarised in Table~\ref{tab:evidences}, complemented by the corresponding Jeffreys scale \citep{jeffreys1998theory, nesseris2013jeffreys} and odds ratio (assuming identical prior model probabilities).

It is apparent from the Bayes factors that it is not possible to distinguish between \lcdm\ and \wcdm\ models using the power spectrum alone, for either \lcdm\ or \wcdm\ ground truth mock data.

In contrast, for the field-level SBI inference it is possible to distinguish between \lcdm\ and \wcdm.  Evidence for the correct underlying ground truth model is nevertheless considered weak on the Jeffreys scale at an odds ratio of 2.10:1 for the \lcdm\ mock data, but it is strong with an odds ratio of 53.5:1 for the \wcdm\ mock data.
Furthermore, mismatches in data and model are now distinguishable visually from the contours of Figure~\ref{fig:sbi_w}, with the incorrect model showing a clear bias. Of course, when analysing real data the true underlying model is not known and so the evidence must be used for model comparison.
The enhanced model constraining power of the field-level SBI analysis due to the extraction of high-order cosmological information is clear.


\subsection{\lcdm\ vs \wowacdm}


\begin{figure*}
	\includegraphics[width=\columnwidth]{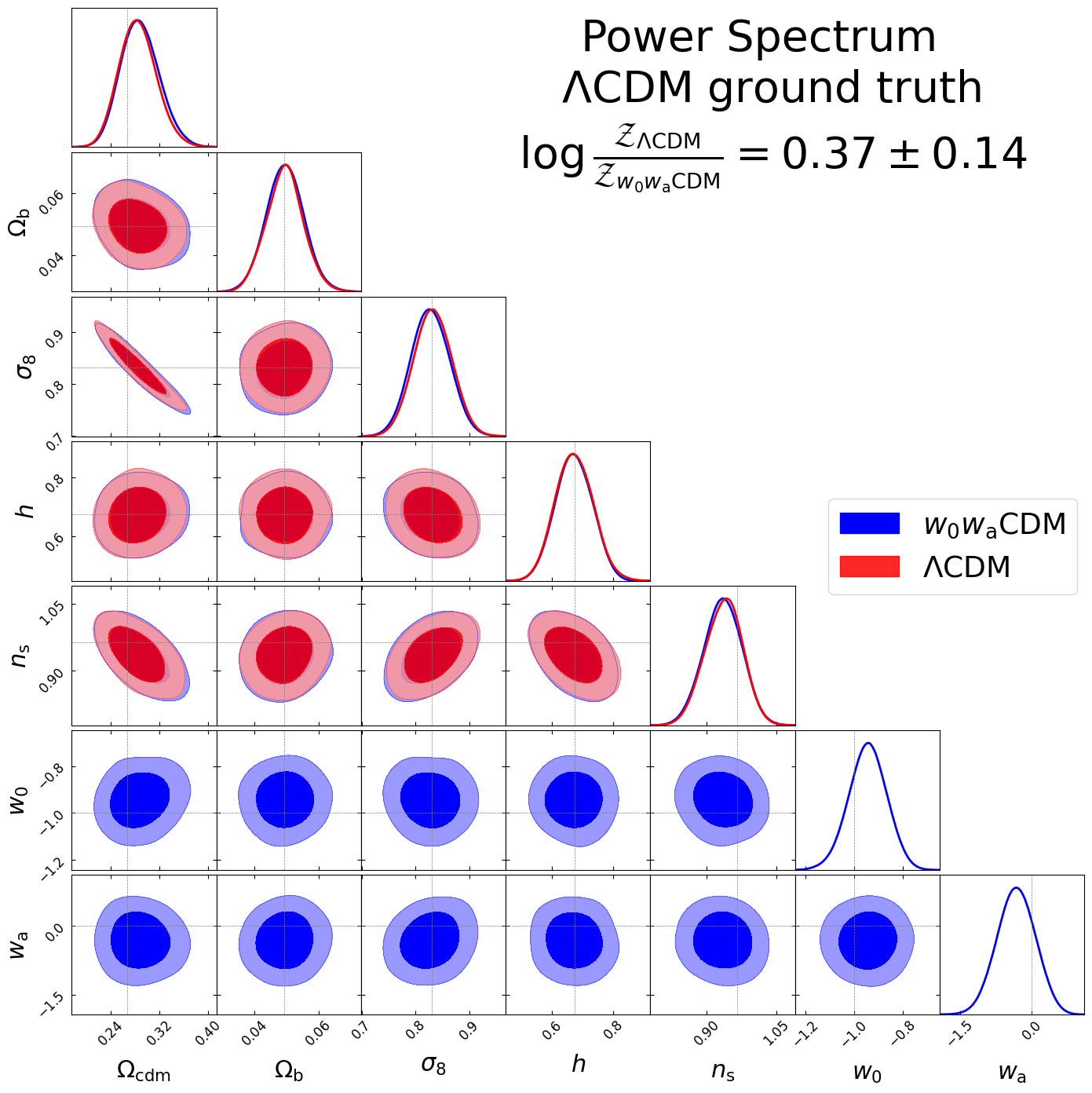}
	\includegraphics[width=\columnwidth]{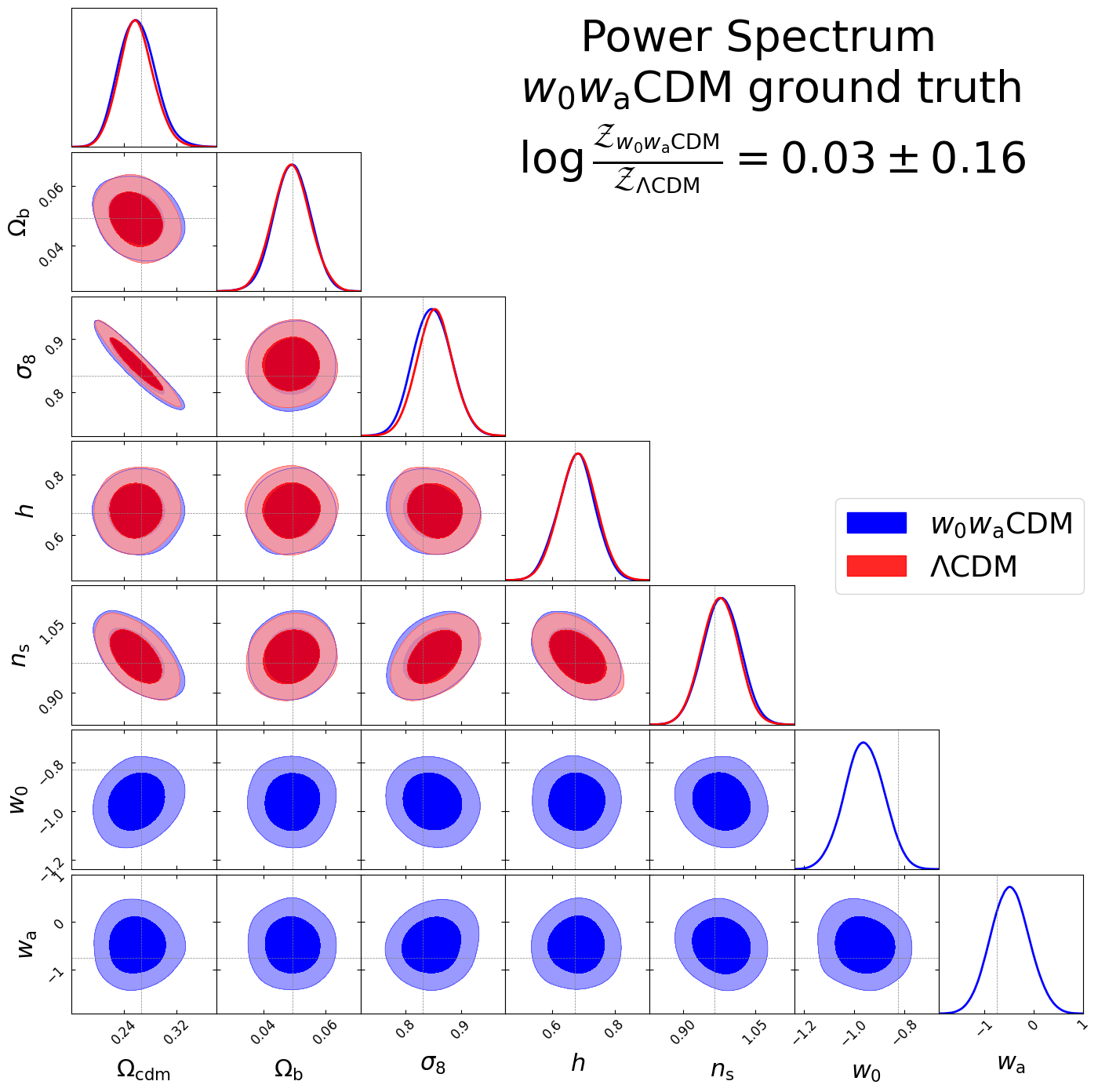}
	\caption{Marginal posterior distributions of cosmological parameters for the \textbf{power spectrum likelihood-based inference}, comparing \textbf{\lcdm\ vs \wowacdm}. Ground truth underlying parameter values are indicated by dashed lines. Left: \lcdm\ ground truth data vector. Right: \wowacdm\ ground truth data vector. For both ground truth scenarios the Bayesian evidence values show it is \textbf{not possibile to distinguish cosmological models}. }
	\label{fig:ps_wowa}
\end{figure*}

\begin{figure*}
	\includegraphics[width=\columnwidth]{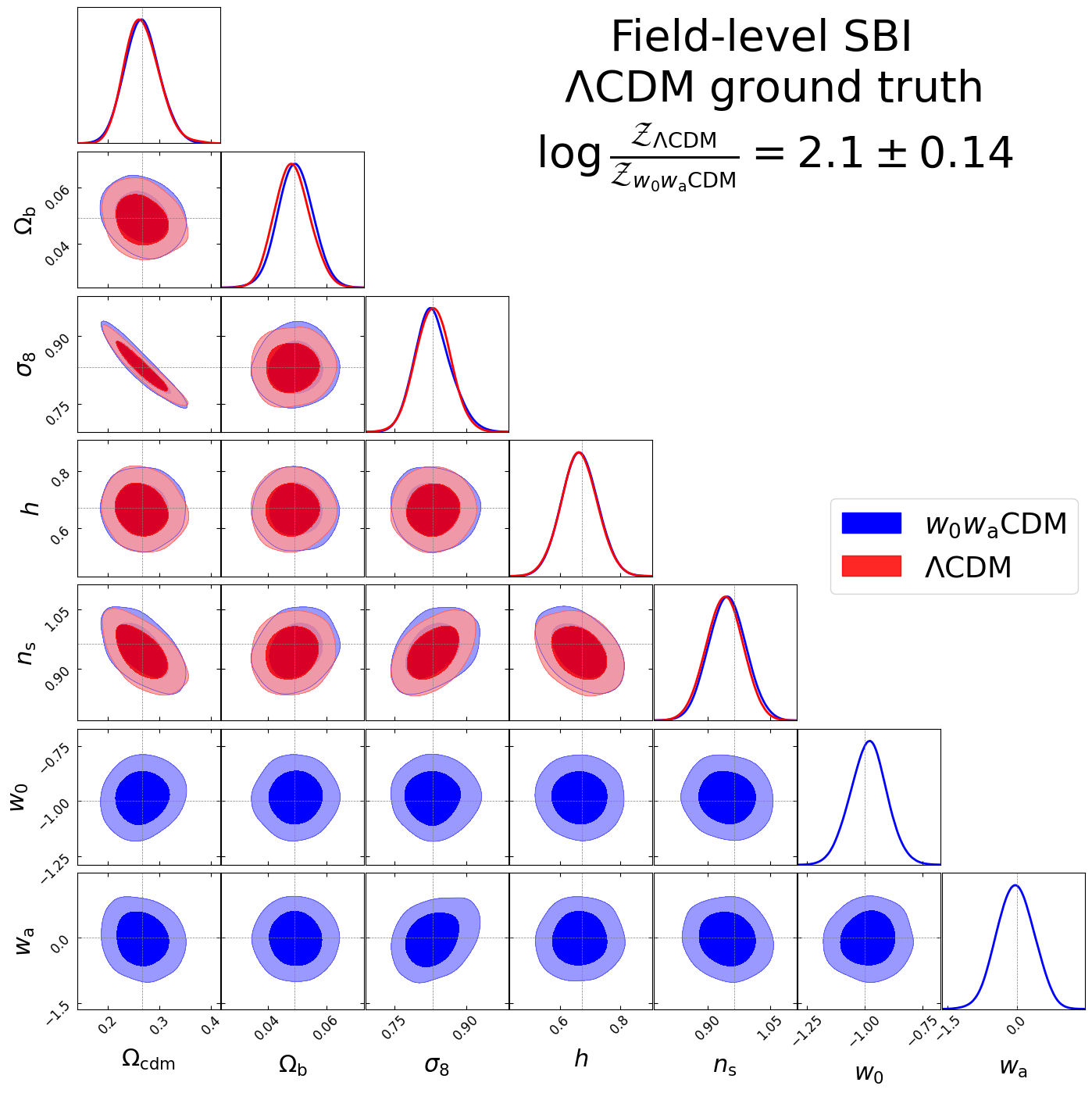}
	\includegraphics[width=\columnwidth]{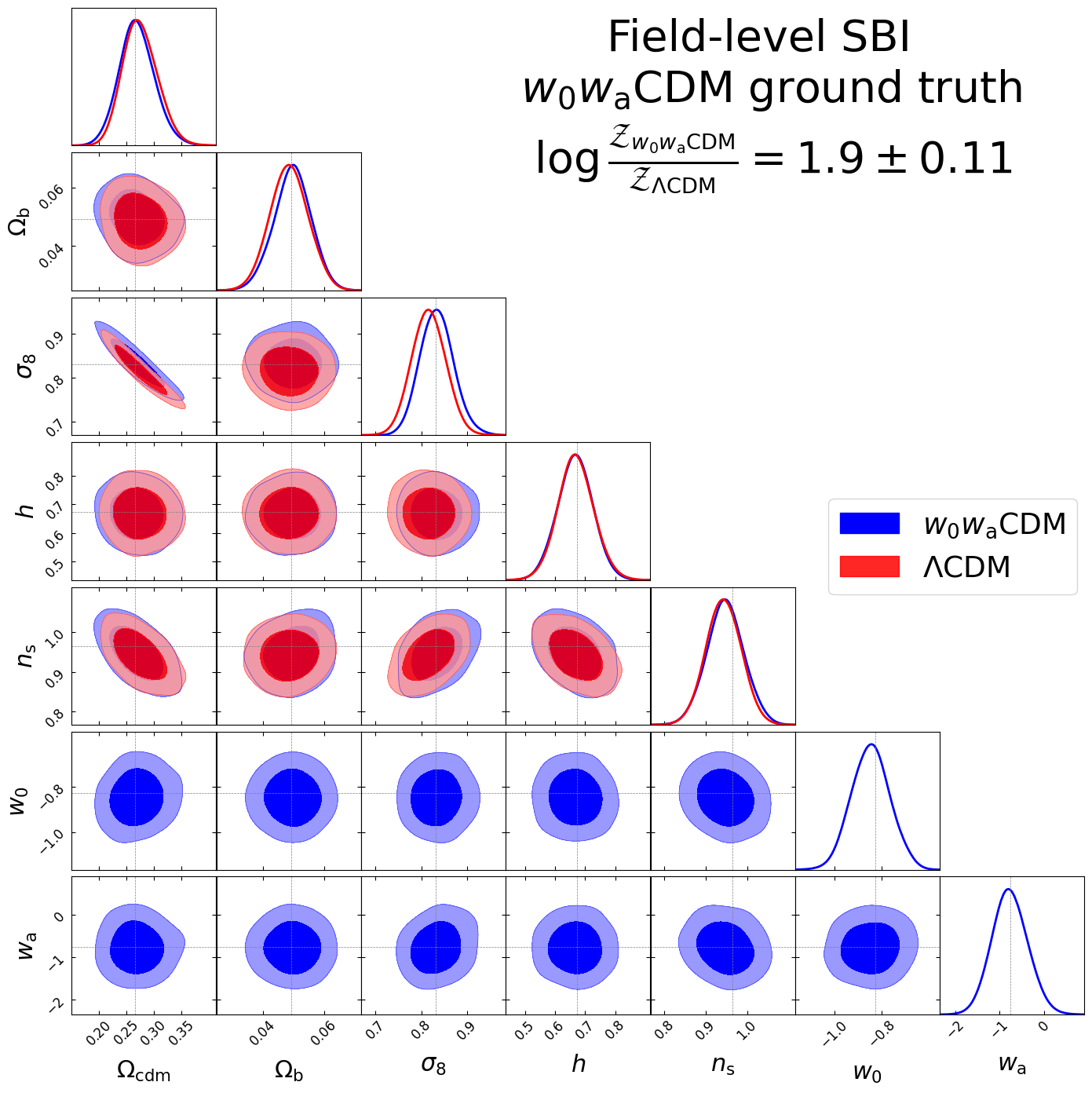}

	\caption{Marginal posterior distributions of cosmological parameters for the \textbf{field-level SBI inference}, comparing \textbf{\lcdm\ vs \wowacdm}. Ground truth underlying parameter values are indicated by dashed lines. Left: \lcdm\ ground truth data vector. Right: \wowacdm\ ground truth data vector. For the former ground truth scenario the Bayesian evidence \textbf{definitively prefers the true underlying model \lcdm}. For the latter ground truth scenario the Bayesian evidence \textbf{definitively pefers the true underlying model \wowacdm}. }
	\label{fig:sbi_wowa}
\end{figure*}

\begin{figure}
	\includegraphics[width=\columnwidth]{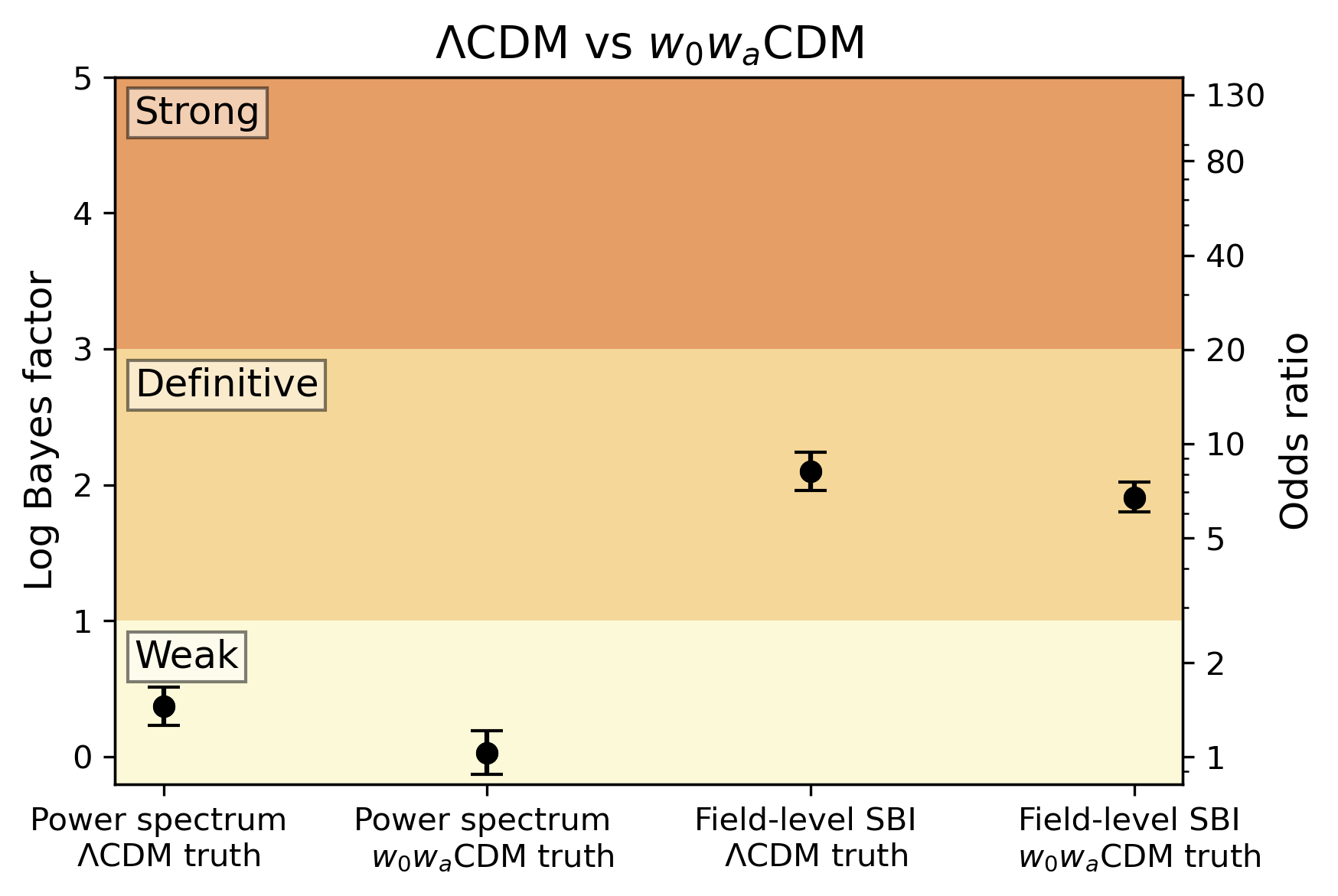}
	\caption{Bayes factors with errors for the \lcdm\ vs \wowacdm\ comparison. The shaded regions correspond to the strength of the Bayes factor on the Jeffreys scale.  Note that \textbf{power spectrum likelihood-based inference cannot distinguish between \lcdm\ and \wowacdm, whereas field-level SBI inference can}. }
	\label{fig:bayes_wowa}
\end{figure}

\begin{table*}
	\caption{Summary of Bayes factors, where they lie on the Jeffreys scale and corresponding odds ratio for the different model comparisons and ground truth data vector model combinations considered. Notably we can see that using \textbf{power spectrum likelihood-based inference cannot distinguish between the cosmological models}, whereas \textbf{field-level SBI inference can distinguish between cosmological models}.}
	\label{tab:evidences}
	\begin{tabular}{cccccc}
		\toprule

		Model              & Ground truth & Method          & Bayes factor (log) & Jeffreys scale & Odds ratio \\
		\midrule
		\lcdm\ vs \wcdm    & \lcdm        & Power spectrum  & $0.08 \pm 0.16$    & inconclusive   & 1.08 : 1   \\
		\lcdm\ vs \wcdm    & \wcdm        & Power spectrum  & $0.16 \pm 0.16$    & inconclusive   & 1.17 : 1   \\
		\lcdm\ vs \wcdm    & \lcdm        & Field-level SBI & $0.74 \pm 0.25$    & weak           & 2.10 : 1   \\
		\lcdm\ vs \wcdm    & \wcdm        & Field-level SBI & $3.98 \pm 0.20$    & strong         & 53.5 : 1   \\ \midrule
		\lcdm\ vs \wowacdm & \lcdm        & Power spectrum  & $0.37 \pm 0.14$    & inconclusive   & 1.45 : 1   \\
		\lcdm\ vs \wowacdm & \wowacdm     & Power spectrum  & $0.03 \pm 0.16$    & inconclusive   & 1.03 : 1   \\
		\lcdm\ vs \wowacdm & \lcdm        & Field-level SBI & $2.10 \pm 0.14$    & definitive     & 8.17 : 1   \\

		\lcdm\ vs \wowacdm & \wowacdm     & Field-level SBI & $1.90 \pm 0.11$    & definitive     & 6.69 : 1   \\

		\bottomrule
	\end{tabular}
\end{table*}


Figure~\ref{fig:ps_wowa} and Figure~\ref{fig:sbi_wowa} show marginalised posterior distributions recovered for the power spectrum likelihood-based inference and field-level SBI inference, respectively, for \lcdm\ vs \wowacdm. Both figures show results for the two different ground truth mock data cases.
Bayes factors for each setting are displayed on each marginal distribution plot.  Furthermore, they are summarised visually in Figure~\ref{fig:bayes_wowa} and also included
in Table~\ref{tab:evidences}.

Similar to when comparing \lcdm\ and \wcdm, it is apparent from the Bayes factors that it is not possible to distinguish between \lcdm\ and \wowacdm\ models using the power spectrum alone, for either \lcdm\ or \wowacdm\ ground truth mock data.

In contrast, for the field-level SBI inference it is possible to distinguish between \lcdm\ and \wowacdm\ and the correct underlying model is selected.  On the Jeffreys scale model selection is definitive for both ground truth mock data scenarios.  It should also be noted that posterior contours are also more accurately centred on the ground truth for the field-level SBI inference, unlike for the power spectrum inference.

\section{\label{sec:conclusions}Conclusions}
We present a framework that for the first time allows Bayesian model comparison to be performed for field-level inference of cosmological models.  We achieve this by leveraging SBI so that a reduced parameter space containing only cosmological parameters of interest need be considered.
This reduces the dimensionality of the parameter space considerably compared to likelihood-based field-level inference where the pixels of the initial or observed field are treated as parameters to be inferred.  Specifically, we take an NLE (neural likelihood estimation) approach, training a density estimator to learn a surrogate for the likelihood, using the \texttt{CosmoPower} emulator \citep{CP,CPJ} to accelerate the generation of simulations needed for training.  We then perform MCMC sampling to generate posterior samples, which are not only used for parameter estimator but to also compute the Bayesian evidence for model selection using the learned harmonic mean estimator implemented in the \texttt{harmonic} code \citep{mcewen2021machine, polanska2024learned, spurio2023bayesian}.

We apply our framework to mock Stage IV cosmic shear observations to assess its effectiveness at distinguishing between various models of dark energy.  For comparison purposes we also consider a traditional power spectrum likelihood-based inference.  Due to the higher order statistical information extracted by our field-level SBI approach, it is able to successfully distinguish dynamical dark energy from \lcdm\ using the Bayesian evidence, whereas the power spectrum inference cannot.
If the DESI results that hinted at the possibility of dynamical dark energy \citep{adame2024desi} were indeed the true underlying model, Stage IV surveys such as those by Euclid and Rubin-LSST, would be able to provide definitive evidence for dynamical dark energy.

Given the effectiveness of our field-level cosmological model selection framework, it is important to extend it to a more realistic setting in preparation for application to Stage IV surveys.  In particular, more realistic simulations, observational effects and systematics need to be incorporated in the forward model.  Furthermore, the forward modelling, any field-level emulation, and compression must be extended to the spherical setting to support the wide fields of upcoming surveys (\textit{e.g.} using spherical machine learning or scattering techniques; \cite{cobb:efficient_generalized_s2cnn, ocampo:disco, mcewen:scattering, mousset:spherical_scattering_covariance}). Field-level SBI techniques exhibit significant promise and when applied to upcoming Stage IV surveys could provide one of the most effective means of determining the underlying nature of dark energy.


\begin{acknowledgements}
	KL is supported by STFC (grant number ST/W001136/1). JDM is supported by EPSRC (grant number EP/W007673/1) and STFC (grant number ST/W001136/1).
\end{acknowledgements}


\bibliography{bibliography}



\appendix
\section{Configuration}\label{app:config}

While the general methodology is described in Section~\ref{sec:methods}, we outline here the specific settings and parameters configured to produce the results presented in Section~\ref{sec:results}.

\subsection{Survey settings}

To simulate mock Stage IV survey data we follow the Rubin-LSST science requirements document \citep{lsst2018desc} and target a Y10 data release. This means that the underlying source galaxy redshift distribution follows a Smail distribution \citep{smaill1995deep} parameterised by
\begin{equation}
	n(z) \propto z^2 \exp \left(-\frac{z}{z_0}\right)^{\alpha},
\end{equation}
\noindent with $z_0 = 0.11$ and $\alpha = 0.68$ and with 5 redshift bins each containing an equal number of galaxies and photometric redshift error given by $\sigma_z = 0.05(1+z)$. To model survey observational noise, we assume a shape noise of $\sigma_e = 0.26$ and a galaxy number density of $n_g = 27$ arcmin$^{-2}$. Following Ref. \cite{lanzieri2024optimal} for \href{https://github.com/DifferentiableUniverseInitiative/sbi_lens}{\texttt{sbi\_lens}} we set the pixel area $A_{\rm pix} = 5.49$ arcmin$^2$ and observed area to $10 \times 10$ deg$^2$. As such, we model survey noise as additive Gaussian noise with zero mean and variance per tomographic bin given by
\begin{equation}
	\sigma_{\rm noise}^2 = \frac{\sigma_e^2}{n_g A_{\rm pix}}.
\end{equation}

\subsection{Compression}

Compression is performed with a convolutional neural network with a ResNet-18 architecture \citep{he2016deep}. The network is implemented in \href{https://github.com/google-deepmind/dm-haiku}{\texttt{Haiku}} \citep{hennigan2020haiku}. Following Ref.~\cite{lanzieri2024optimal}, we make use of a Variational Mutual Information Maximization (VMIM) loss function introduced to cosmology in Ref.~\cite{jeffrey2021likelihood} which is shown to produce sufficient statistics for SBI.
We train our own compression with the aforementioned architecture following the same procedure described in \citep{zeghal2024simulation} and included in \href{https://github.com/DifferentiableUniverseInitiative/sbi_lens}{\texttt{sbi\_lens}}.

\subsection{SBI NLE density estimator}

We make use of a masked autoregressive flow (MAF) \citep{papamakarios2017masked} as the conditional neural density estimator. The MAF is constructed out of 5 masked autoencoders for density estimation (MADE) \citep{germain2015made} with 50 hidden features each. The NLE density estimator is trained using the \href{https://github.com/sbi-dev/sbi}{\texttt{sbi}} software package \citep{tejero-cantero2020sbi}. Following the work of Ref.~\cite{lanzieri2024optimal} we make use of 150,000 compressed simulations for training, which is likely more than strictly necessary.

\subsection{MCMC}

To obtain posterior samples, we make use of NUTS \citep{Hoffman14} implemented in \href{https://github.com/pyro-ppl/numpyro}{\texttt{NumPyro}} for the power spectrum analysis as detailed in Section~\ref{sec:methods_ps} and \href{https://github.com/dfm/emcee}{\texttt{emcee}} \citep{foreman2013emcee} for the field-level SBI analysis as detailed in Section~\ref{sec:methods_field}. For NUTS we set the number of chains to 3, with a burn-in length of 1200 and chain length of 1800. For \href{https://github.com/dfm/emcee}{\texttt{emcee}} we run 24 walkers with 200 burn-in steps and 300 samples per walker. In both cases this results in 7200 samples after burn-in. We plot our contours with the \href{https://github.com/cmbant/getdist}{\texttt{getdist}}\software{https://github.com/cmbant/getdist} software package \citep{lewis2019getdist}.
\subsection{Learned harmonic mean estimator}

For the internal learned target distribution of the learned harmonic mean we train a rational quadratic spline flow \citep{durkan2019neural}, including standardization \citep{polanska2024learned}, consisting of $2$ layers, with $128$ spline bins.  For the results presented in this article we concentrated the flow using a temperature of $T=0.8$, although we also found overall results were robust to changing the temperature from 0.4 to 0.9 in steps of 0.1.  Of the available MCMC samples, 50\% were used for training the flow and 50\% for evidence calculation.


\end{document}